\documentstyle[preprint,revtex]{aps}
\begin{document}
\begin{title}
\begin{center}
Area Decay Law Implementation for Quark String Fragmentation
\end{center}
\end{title}

\author{E.G.Gurvich \cite{gur} and G.G.Leptoukh \cite{lep}}

\begin{instit}
Institute of Physics, 6 Tamarashvili st., Tbilisi--77, Republic of Georgia
\end{instit}

\begin{abstract}

We apply the Area Decay Law (ADL) straightforwardly to simulate a quark
string hadronization and compare the results with the explicit analytic
calculations.

We show that the usual "inclusive" Monte--Carlo simulations do not correspond
to the ADL because of two mistakes: not proper simulation of
two--dimensional probability density and lack of an important
combinatorial factor in a binary tree simulation. We also show how to
simulate area decay law "inclusively" avoiding the above--mentioned
mistakes.  \end{abstract}

\section{Introduction}

Models of string (flux tube) soft hadronization serve as
convenient tools for the multiple hadron production computation. The so
called quark strings (with quark and antiquark at the string ends) are
produced in hadron and lepton scattering processes after the
interaction.
We are interested here in strings with endpoint quarks almost on mass
shell, where the QCD cascade problems are hidden within the string
itself or are absent at all.

An Area Decay Law (ADL) assumes a constant probability $w$ for a quark
string break to occur in an element of invariant two--dimensional
area $dW = w \cdot
dA$, where $dA$ is an element of space--time area sampled by a string.

One can find a probability that no breaks occur in the
invariant area $A$ under consideration \cite{artru,cnn}:

\begin{equation}
P(A) = e^{-w \cdot A}.
\end{equation}

In a certain one--dimensional $\tau$ evolution it is possible to find a
differential probability that the {\it first} break occurs after an
invariant area $A(\tau)$ is swept out:

\begin{equation}
dw(A) = w \ e^{-w \cdot A} dA,
\end{equation}
that surely has much in common with the well--known radioactive decay law.

There are several ways of string fragmentation simulation.  In the
so--called "exclusive" approach break points are simulated step
by step according to the certain parameter evolution.  It can be
done either from the one string end toward another one (like in the
old LUND scheme {\cite{lund}), or in the real time order by generating
breaks according to their real space--time development in a fixed frame
of reference as in CALTECH--II\cite{mor}, etc.

In an "inclusive" break simulation (VENUS \cite{ven}) an {\it
arbitrary} break point is simulated. Then each of two appearing string
fragments are treated independently of each other with an {\it
arbitrary} break simulation of the same kind.

We've found out that, as a rule, in an "inclusive" simulation two
"hidden" assumptions are made:

i) the ADL formula (2) is applied to an {\it arbitrary} break
instead of the {\it first} break probability computation.

ii) an important combinatorial factor related to a binary tree structure
of a simulated cascade is omitted.

These two mistakes lead to a rapid development of hadronization cascade
causing the high multiplicity and a strong distortion of
inclusive spectra.

In the present article we show how to simulate quark string hadronization
properly both "exclusively" and "inclusively".
To make our analysis more transparent we focus on a
consideration of the so called primary string fragments (clusters,
resonances etc. following the known notation \cite{ff}), but not on the
final hadrons, because this allows us to compare the MC results with
the direct analytic calculations. To simplify our analysis we deal here
with massless quarks of a single flavor only.  We work here under the
condition that the primary fragment is formed when a quark and an
antiquark it consists of meet each other at a certain point. We assume
that the fragment then evolves the known "yo-yo" evolution without
additional string--like breaks. This is, up to our knowledge, the most
natural termination of string fragmentation cascade \cite{lund}.

In Sec.II we present an analytic calculation of fragment rapidity spectrum
to be compared with the Monte--Carlo simulations;
in Sec.III we recall the rule for two-dimensional probability
simulation and apply this rule to "exclusive" and "inclusive" simulations;
discussion and conclusion are made in Sec.IV.

\section{Analytic calculations}

As always, a string endpoint quark motion is determined by a simple
Newtonian rule: during a period of time $t$ a quark momentum changes by
$\kappa \cdot t$, where $\kappa \approx 1 GeV / fermi$ is a quark string
tension.
Putting for convenience $\kappa = 1$ we can work in $t,x$ variables
or in the "light--cone" variables:
$x^+= (t + x) / \sqrt{2}$ and $x^- = (t - x) /\sqrt{2}$.
In the latter case we rotate the well--known picture of string breaking
evolution by $45$ degrees for better visualization.

For all the calculations below we use the following recipe:  to
calculate a certain spectrum  one has to fix the number of
break--points required for this final state and integrate over all
available phase space. The only limitation is to guarantee the absence
of additional breaks in the absolute past of this fixed state.

To obtain primary fragment inclusive spectrum we have to
choose an arbitrary fragment (see \cite{gl}) and calculate a
probability for such a fragment to have a rapidity in an interval $(y,
y + dy)$ irrespective of all other fragment behavior. Two distinct
cases are to be considered here: in the central case quark and
antiquark from two neighboring break points are chosen to create a
fragment, while in the edge case --- at least one of string endpoint
quarks constitutes a fragment. For simplicity, a three
fragment event is presented in Fig.1, where the numbers $1$ and $2$
denote break--points with the light--cone coordinates $(x^+_1,x^-_1)$
and $(x^+_2,x^-_2)$ correspondingly, while capital letters denote
areas to be used in the ADL.

For the central case spectrum  the following is written:

\begin{equation}
f_c(y) = w^2 \int_0^{X^-} dx^-_1 \int_0^{x^-_1} dx^-_2 \int_0^{X^+}
dx^+_2 \int_0^{x^+_2} dx^+_1 e^{-wx^-_1x^+_2} \delta (y - {1\over 2}
\ln{{x^+_2 - x^+_1} \over {x^-_1 - x^-_2}}),  \end{equation}

The exponent in
the integrand guarantees the very existence of the central fragment,
i.e., that no other breaks occur in the absolute past of the meeting--point
under consideration (a rectangular area in Fig.1 determined by the area
$B + D + C + E$).  The Dirak $\delta$--function fixes the
break--point rapidity.

For the right edge case spectrum ($C + E + F$) the expression is the following:

\begin{equation}
f_R(y) = w\int_0^{X^-} dx^-_2 \int_0^{X^+} dx^+_2 e^{-wx^-_2X^+}
\delta (y - {1\over 2} \ln{{X^+ - x^+_2} \over x^-_2}).
\end{equation}

A similar expression can be written for the left edge ($A + B + C$)
with the obvious substitutions.

To complete the consideration we have to add a small contribution
from the case of a string having no breaks at all during its evolution:

\begin{equation}
f_0(y) = e^{-wX^-X^+}\delta (y - {1 \over 2} \ln{X^+ \over X^-}).
\end{equation}

Summing up all the contributions we obtain the following expression
for $y > 0$ in the CMS (putting $X^+ = X^- = \sqrt{s/2}$):

\begin{equation}
f(y) = 1 + {2 \over {ws}} \left \{ e^{2y}
\left [1 - e^{-0.5\cdot {ws} \cdot \exp{(-2y)}} (1 + ws e^{-2y}) \right
] + e^{-2y} \left [1 - e^{-0.5\cdot {ws}} (1 + ws) \right ] \right \}.
\end{equation}

In Fig.2 the above--mentioned contributions to the primary fragment
spectrum are shown explicitly.

Some interesting aspects of this spectrum properties such as energy
behavior and "humps" at the edges will be considered in a separate
publication.

Integrating the spectrum over the whole rapidity interval we obtain the
well--known fragment mean multiplicity energy behavior:

\begin{equation}
\langle n \rangle \approx \ln{{w s \over 2}} + C + 1, \end{equation}
where $C$ is the Euler constant ($C = 0.5773$). Here $s = 2 A$ is the
CMS energy squared.

\section{Simulation}
Naturally, when working with the ADL a
space--time string area is a convenient coordinate to work in. Below
we show that a choice of the conjugated coordinate is of importance.

In fact, let us recall the known rules for the Bayes theorem based probability
simulation. When dealing with two-dimensional probability, say,
$p(a,\xi)$, one has to do the following:

1. to construct an {\it unconditional}
marginal probability density for $a$ variable simulation:
\begin{equation}
p(a) = \int d\xi \ p(a,\xi).
\label{s1}
\end{equation}

2. to calculate a {\it conditional} probability density for $y$
variable simulation:
\begin{equation}
p(\xi | a) = p(a,\xi) / p(a).
\label{s2}
\end{equation}

To deal with one--dimensional ADL expression (2) ({\it a la}
radioactive decay law) it is necessary to have $p(a) = p(a,\xi)$.
But from (\ref{s1}) it follows that this is not the general case.

\subsection{"Exclusive" simulation}

In an "exclusive" approach we work in coordinates that allow us to use
the one--dimensional law (2) directly. We present here a simplest
version that deals with a single fragment to be broken at each step.

We go from the light--cone variables  $x^+$ and $x^-$ to the more appropriate
variables $\hat{a} = x^+ X^-$ and $\hat{x^-} = x^- / X^-$ (the
Jacobian is equal to unity).  To illustrate this $\hat{a}$--evolution we
again use Fig.1. For the first step $\hat{a}= A + B + C$, for the second one
$\hat{a}= D + E$ and for the last step $\hat{a}= E$.

Here the formulae (\ref{s1}) and (\ref{s2}) required for the
simulation of $p(\hat{a},\hat{x^-}) = w \cdot exp{(-w\hat a)}$
should be written as:
\begin{eqnarray}
p(\hat{a}) = w \int_0^1 d\hat{x^-}\ p(\hat{a},\hat{x^-}) = w e^{-w
\hat{a}}, \qquad 0 < \hat{a} < A.\nonumber\\
p(\hat{x^-} | \hat{a}) = 1, \qquad 0 < \hat{x^-} < 1.  \end{eqnarray}

Thus, according to the ADL (2), we compute $\hat{a}$; if it is less
than the whole allowed area $A = X^- X^+$, then the string can be broken down.
Knowing $\hat{a}$ we then simulate $x^-$ randomly along $x^+=const$
line ($x^+$ is calculated from $\hat{a}$).

For the next break--point simulation (the first one
for the next fragment!) the same rule is applied etc.

The most attractive feature of this evolution is that at each
step we have a single string to be fragmented. This
is to be compared with, for example, the real--time evolution
\cite{mor}, where each break provides for two fragments to be broken
later.

This simulation remarkably agrees (Fig.2) with the results of the
corresponding analytic calculations for fragment inclusive
spectrum.

\subsection{"Inclusive" simulation}

The usual "inclusive" simulation deals with an arbitrary break--point
simulation \cite{ven}. At each step two fragments are produced. Then
each of these fragments are broken into two new ones and so on ---
this is a typical binary tree. This approach is very convenient for the
Monte--Carlo implementation.

The conjugated variables are $a$ and $y$,
where $a=x^+ x^-$ is the area sampled by a string in the absolute past
of the break, while $y = \ln{(x^+ / x^-)}$ is the break--point rapidity.
Again the Jacobian is equal to unity, and the ADL's exponent is
simply $p(a,y) = w \cdot exp{(-w a)}$:
\cite{ven}.

1. However, according to (\ref{s1}) and (\ref{s2}) we should write for
$p(a,y)$ simulation the following expressions:
\begin{eqnarray} p(a) =
w \int_{y_{min}}^{y_{max}} dy\ p(a,y) =w e^{-w {a}} \ln{ A \over a},
\qquad 0 < a < A. \nonumber\\ p(y | a) = {1 \over \ln{ A \over a}},
\qquad y_{min} < y < y_{max}.  \end{eqnarray} where $y_{min} =
\ln{(\sqrt{a} / X^-)}$ and $y_{max} = \ln{(X^+/\sqrt{a})}$ are limits
for $y$ at a fixed value of $a$.

We see that the $a$ area should not be simulated by a single exponent
law as it has been done, for example, in the VENUS Monte--Carlo code
\cite{ven}.

Here the notation "inclusive" becomes more transparent. In fact, since
an {\it arbitrary} break--point is considered, integrating $p(a)$ in
(11) over $a$ we obtain the mean number of breaks in the allowed area
$A$ that is evidently equal to $\langle n \rangle -1$ from (7).
Contrary to this, an expression in (10) for the "exclusive" approach is
integrated to the quantity equal to a number of the {\it first} breaks
in the area $A$, which is evidently less than unity.

2. Another important point is connected with a multiple counting in
an "inclusive" simulation due to a binary tree structure of simulated
cascades.

   In fact, an event of the certain final configuration (break point
number and coordinates) has a certain "history" --- the exact sequence
of fragment generations.

   In an "exclusive" simulation we always have the unique "history"
of an event - there is one--to--one correspondence between an event
and its tree. Note, that the same event for a different evolution
parameter can correspond to a quite another topology (but again to a
single tree!).

   Contrary to this, in an "inclusive" approach an arbitrary break
simulation leads to $W_n$ topologically nonequivalent trees for event
with the certain final configuration with $n$ breaks.

   We can illustrate this by considering the event in Fig.1.  It is
easy to see that this event with two string breaks can be obtained in
two different ways:  the first "inclusive" break being either at the
point 1 ( $\exp{\left (-w*(B + C) \right )}$ in the ADL) or at the
point 2 ($\exp{\left (-w*(C + E)\right )}$ in the ADL). It means that
the event shown in Fig.1 is counted twice in an "inclusive" simulation
while being counted once only in any "exclusive" approach.

   For an event with three breaks we can draw 5 nonequivalent
"histories" giving the same event. So during "inclusive" simulation
using different sets of random values we obtain 5 identical events,
thus increasing the weight of "3--break" events by the factor 5, and
so on.

   The number of "histories" or nonequivalent trees giving an event
with $n$ breaks is well--known (ref.\cite{bin}) --- it is the number
of binary trees with $n$ knots:

\begin{equation}
W_n = {1 \over {n+1}}{2n\choose {n}}, \qquad W_0=W_1=1, \qquad W_2=2, W_3=5,
\ldots
\end{equation}

Therefore, we have shown that each event is multiple counted in an
"inclusive" simulation evidently leading to higher multiplicity with
the other obvious consequences.

\subsection{Corrected "inclusive" simulation}

Keeping in mind all the above--mentioned factors we propose a
corrected "inclusive" version of a quark string hadronization.

1. to simulate an {\it arbitrary} break one should use $exp(-w a)
\cdot \ln{(A/a)}$ instead of $exp(-w a)$. This can be done either by
the direct simulation of (11) or by the weight techniques we have
used in the present consideration. Following \cite{ven} we
simulate $a$ according to $exp(-w a)$,  but then weight each event
by the factor equal to $\ln{(A/a)}$ weight product from all the breaks
in the event.

2. Each generated event with $n$ breaks we weight by
$1 / W_n$ (see the previous chapter).

Taking into account the above--mentioned weights we obtain the same result
in the corrected "inclusive" simulation as in analytic
calculations and in the "exclusive" simulation.  In Fig.3  we present
the primary fragment inclusive spectrum computed by various techniques.

\section{Discussion and conclusions}

To check our observation we have tested the real time evolution
simulation used in \cite{mor} (also an "exclusive" one due
to our definition). Surely, a binary tree is created in this
evolution. However, each break is labeled here by the time variable,
and the possible double counting is avoided.

In our "exclusive" simulation we do not encounter a problem
with combinatorics, as we always have a single "history" corresponding to
event with $n$ breaks and the exact sequence of breaks is completely defined.
It looks like the "salami" structure of
a simulation as in the LUND scheme \cite{lund}, where fragments or hadrons
are fallen from a string ends only, giving a single topology per event.

Contrary to this, in a binary tree "inclusive" simulation one
has $W_n$ topologically nonequivalent trees for event with $n$ breaks.

Being unaware of the above--mentioned mistakes in the "inclusive"
approach
one is forced to use an additional cascade damping to obtain
the reasonable multiplicity etc. For example, string--like breaking of
fragments with masses greater than a certain threshold value \cite{ven}
has been forbidden.
In fact, it reduces the multiplicity, the same time distorting other
important properties such as correlation etc. This is because a
mixture of events of various multiplicities causes a nontrivial
long--range correlations that in reality are absent in a single quark
string \cite{egg}.

To conclude with, we state that there are two important points
to be taken into account in any "inclusive" simulation
of a string breaking. The first one is connected with a more
careful use of the Area Decay Law applied to an {\it arbitrary}
break simulation, while the second one follows from the
binary tree structure of such a cascade simulation.

\nonum
\section{Acknowledgments}

We wish to acknowledge our gratitude to V.Abramovsky and Yu.Werbetsky
for a valuable help.

The work was supported, in part, by a Soros Foundation Grant awarded
by the American Physical Society.

\newpage
\figure{The light--cone string fragmentation picture:
numbers $1$ and $2$ denote string break--points,
capital letters denote areas used for the ADL simulation.}
\figure{Analytic calculations vs. "exclusive" MC simulation. Lines correspond
to different contributions to fragment rapidity spectrum, while the
simulation results are shown by circles.}
\figure{Primary fragment inclusive spectrum for various simulations.}

\end{document}